# CHAOS BASED MIXED KEYSTREAM GENERATION FOR VOICE DATA ENCRYPTION

Musheer Ahmad[1], Bashir Alam[1] and Omar Farooq[2]

[1]Department of Computer Engineering, Jamia Millia Islamia, New Delhi, India
musheer.cse@gmail.com, babashiralam@gmail.com
[2]Department of Electronics Engineering, ZH College of Engineering and Technology, AMU, Aligarh, India
omar.farooq@amu.ac.in

## ABSTRACT

*In this paper, a high dimensional chaotic systems based mixed keystream generator is proposed to secure the voice data. As the voice-based communication becomes extensively vital in the application areas of military, voice over IP, voice-conferencing, phone banking, news telecasting etc. It greatly demands to preserve sensitive voice signals from the unauthorized listening and illegal usage over shared/open networks. To address the need, the designed keystream generator employed to work as a symmetric encryption technique to protect voice bitstreams over insecure transmission channel. The generator utilizes the features of high dimensional chaos like Lorenz and Chen systems to generate highly unpredictable and random-like sequences. The encryption keystream is dynamically extracted from the pre-treated chaotic mixed sequences, which are then applied to mask the voice bitstream for integrity protection of voice data. The experimental analyses like auto-correlation, signal distribution, parameter-residual deviation, key space and key-sensitivity demonstrate the effectiveness of the proposed technique for secure voice communication.*

## 1. INTRODUCTION

With the advancement of modern wireless telecommunication and multimedia technologies, a huge amount of sensitive voice data travels over the open and shared networks. Voice-based communication becomes prominent in the application areas of military, voice over IP, e-learning, voice-conferencing, phone banking, phone stock market services, news telecasting etc. These applications are critical with respect to integrity protection of voice data and privacy protection of authorized users. The probable security threats in a voice-based communication system as highlighted by voice over IP security alliance [1] are: social threats, interception and modification threats, denial of service threats, service abuse threats, physical access threats and interruption of service threats. Hence, the need of high level security system is pre-requisite of any secure voice communication system to forestall these attacks. The cryptographic techniques are to be developed and deployed which can address and fulfil the increasing security demands of secure voice-based communication. The conventional cryptographic techniques are efficient for the text data. But they computationally fail in providing ample security due to the bulk data capacity and high redundancy of voice data. Therefore, the design of efficient voice security methods demands new challenges which can provide high security to the voice data. To achieve this, a number of voice encryption techniques have been suggested [2-12]. Among them, the chaos-based techniques are considered efficient for dealing with bulky, redundant voice data. They provide fast and highly secure encryption methods. This is because of the reason that the chaotic systems are characterized with high sensitivity to its initial conditions, ergodicity, random behaviour, and long periodicity. The cryptographic properties such as diffusion, confusion and disorder can be achieved by applying iteration operations to these systems.

The challenging task in the design of cryptographic techniques is to generate keystreams of high randomness and statistical properties. The quality of the keystream generated by the security system determines its strength from cryptographic viewpoint. The importance of a careful design of cryptographic keystream generators cannot be underestimated as these generators are becoming particularly useful to ensure secure multimedia data transmission over an insecure communication channels. Generating keystreams with high randomness is a vital part of many cryptographic operations. A high quality keystream generated by a cryptographic system is characterized by the following main properties [13].

*Randomness*: It passes most common standard randomness tests and should have good statistical properties.

*Repeatability*: It gives the same output sequence when the same seed is used.

*Unpredictability*: If the seed is unknown, the next output bit in the sequence is unpredictable in spite of any knowledge of previous bits in the sequence.

*Long Period*: The deterministic algorithm that generates the pseudo-random sequence has a fixed period that must be as long as possible.

A chaos-based keystream generator is proposed for symmetric voice data encryption to meet the demands of high security, privacy and reliability of secure voice communication system. The features of high dimensional chaotic systems are exploited in the design. The sequences generated by the 3D chaotic systems are pre-processed, quantized and then mixed to produce cryptographically and statistically better keystream, which is applied to encrypt the voice data to evaluate its encryption performance. The results support the effectiveness and suitability of the proposed design for voice data encryption.

## 2. PROPOSED KEYSTREAM GENERATION

The one dimensional chaotic systems have some inherent weaknesses such as: (1) they provide low key space, (2) their iteration operations generate single sequence and (3) they are weak against adaptive parameter synchronous attack [14]. Therefore, the high-dimensional Lorenz and Chen chaotic systems are employed in the design. Each of these systems generates three distinct stochastic chaotic sequences on iteration operations, which makes encryption faster. Moreover, the Lorenz and Chen systems are more complex and generate more unpredictable sequences than one-dimensional chaotic systems.

The Lorenz chaotic system is described by the following differential equations:

$$\begin{aligned} \dot{x}_1 &= \sigma(x_2 - x_1) \\ \dot{x}_2 &= rx_1 - x_1 x_3 - x_2 \\ \dot{x}_3 &= x_1 x_2 - \rho x_3 \end{aligned} \quad (1)$$

Where as, the equations describe the 3D Chen system are given as:

$$\dot{y}_1 = a(y_2 - y_1)$$
$$\dot{y}_2 = (c - a)y_1 - y_1 y_3 + c y_2 \qquad (2)$$
$$\dot{y}_3 = y_1 y_2 - b y_3$$

Where $x_1(0)$, $x_2(0)$, $x_3(0)$ are initial conditions, while $\sigma$, $r$, $\rho$ are positive constants of Lorenz system. Let $\sigma=10$ and $\rho=8/3$, the research shows that the Lorenz system exhibits chaotic behaviour when $r > 24.74$. Where as, $y_1(0)$, $y_2(0)$, $y_3(0)$ are initial conditions and $a$, $b$, $c$ are parameters of Chen system. The Chen system is chaotic for $a=35$, $b=3$, $20 \le c \le 28.4$. The equations of Lorenz and Chen systems are quite similar, but topologically they are very different due to parameters $r$ of Lorenz and $c$ of Chen system. These 3D differential equations are solved using RungeKutta-4 method with step size of 0.001. The ideal cryptographic sequence should have good statistical properties. The pre-processing done in Eq. 3 and 4 enhances the statistical properties of the chaotic sequences generated by the Lorenz and Chen systems [14, 15].

$$\hat{x}_k(i) = x_k(i) \times 10^5 - floor(x_k(i) \times 10^5) \qquad (3)$$

$$\hat{y}_k(i) = y_k(i) \times 10^6 - floor(y_k(i) \times 10^6) \qquad (4)$$

Where $k=1, 2, 3$ and $i > 0$ is iteration count. Now, the pre-processed chaotic sequences $0 < x_k(i)$, $y_k(i) < 1$ are quantized and converted into binary bitstreams $\omega_k(i)$ and $\varphi_k(i)$. The quantization is governed by the following transformation:

$$\omega_k(i) = \begin{cases} 0 & if \quad \hat{x}_k(i) < 0.5 \\ 1 & if \quad \hat{x}_k(i) \ge 0.5 \end{cases} \qquad (5)$$

$$\varphi_k(i) = \begin{cases} 0 & if \quad \hat{y}_k(i) > 0.5 \\ 1 & if \quad \hat{y}_k(i) \le 0.5 \end{cases} \qquad (6)$$

The six bitstreams $\omega_k(i)$ and $\varphi_k(i)$ are combined and mixed using XOR operations according to the rules described in Eq. 7 to generate cryptographically better chaotic mixed bitstreams $\Phi_1$, $\Phi_2$, $\Phi_3$ and $\Phi_4$. On mixing, the bitstreams become highly random and uncorrelated. After mixing operation, the mixed bitstreams are fed to a 4×1 multiplexer which dynamically selects one of randomly generated bits $\Phi_1(i)$, $\Phi_2(i)$, $\Phi_3(i)$, $\Phi_4(i)$ to produce the next member of output keystream. The multiplexer require two select lines $S_1S_0$, the select lines should not be static for dynamic operation of MUX. The select lines are made dependent to the random bits $\omega_k(i)$ and $\varphi_k(i)$ for its dynamic operation. The select lines for iteration $i$ are evaluated as $S_0=\omega_1(i)\oplus\omega_2(i)\oplus\omega_3(i)$ and $S_1=\varphi_1(i)\oplus\varphi_2(i)\oplus\varphi_3(i)$. The diagram of proposed mixed keystream generation process is shown in Fig. 1.

$$\left. \begin{array}{l} \phi_1(i) = \omega_1(i) \oplus \varphi_2(i) \oplus \omega_3(i) \\ \phi_2(i) = \varphi_3(i) \oplus \omega_1(i) \oplus \varphi_1(i) \\ \phi_3(i) = \omega_2(i) \oplus \varphi_1(i) \oplus \varphi_2(i) \\ \phi_4(i) = \omega_3(i) \oplus \omega_2(i) \oplus \varphi_3(i) \end{array} \right\} \qquad (7)$$

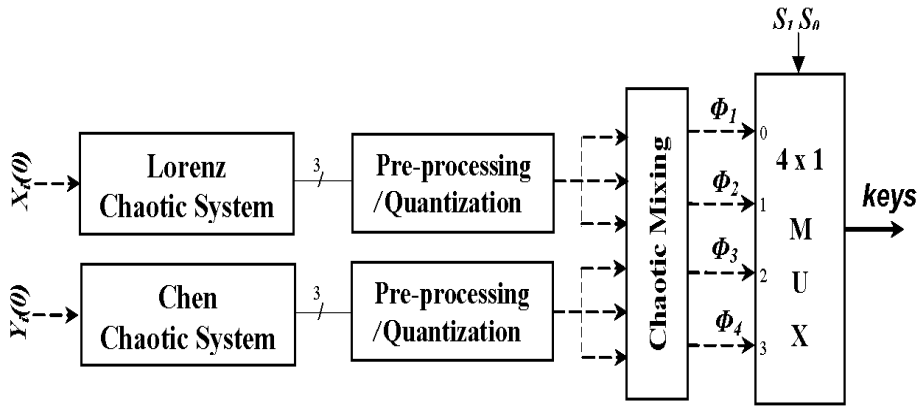

Figure 1. Block diagram of chaos-based mixed keystream generation process.

## 3. RESULTS

In this section, the experimental analyses are presented to demonstrate the effectiveness of the proposed generator. The initial values taken for experimentation are as follows: $X(0)=(x_1(0)=13.3604, x_2(0)=7.2052, x_3(0)=21.5026, \sigma=10, \rho=8/3, r=28)$, $Y(0)=(y_1(0) = -10.058, y_2(0)=0.368, y_3(0)=37.368, a=35\ b=3, c=28)$ and $t=4000$. The two chaotic systems are first iterated $t$ times and these $6 \times t$ values are discarded to remove the transient effect.

The auto-correlation function of the output keystream is shown in Fig. 2. It is clear from the figure that keystream has good delta-function form thereby meeting the requirement of cryptographic random sequence. The function has a maximum value of 0.0056984 for non-zero shift.

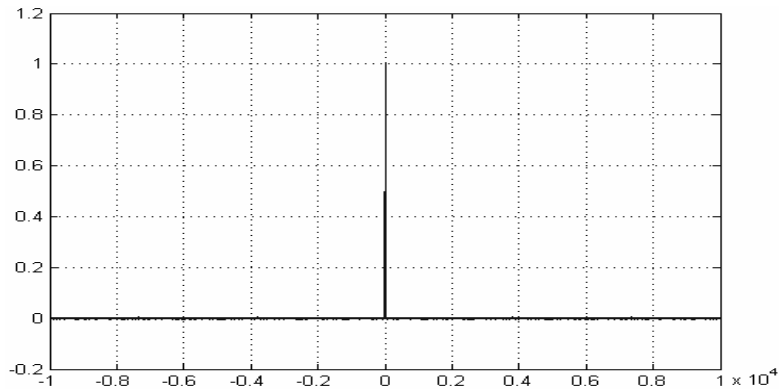

Figure 2. Auto-correlation function of output keystream

### 3.1 Randomness Results

In cryptographic security, suitable metrics are needed to investigate the degree of randomness for binary sequences produced by cryptographic keystream generators. The NIST *Statistical Test Suite* is a statistical package consisting of different tests that are developed to test the randomness of (arbitrarily long) binary sequences produced by either hardware or software based cryptographic systems. The NIST Statistical Test Suite is the result of collaborations between the Computer Security Division and the Statistical Engineering Division at NIST (National Institute of Standards and Technology). The package includes the tests like:

*Frequency*, *Block frequency*, *Cusum-forward*, *Cusum-reverse*, *Runs*, *Longest runs*, *Rank*, *FFT*, *Linear complexity*, *Serial*, *Approximate entropy*, *Lempel-Ziv compression* and *Overlapping template* tests. These tests focus on a variety of different types of non-randomness that could exist in a binary sequence [16]. The generated PN sequences go through the randomness tests.

The NIST randomness tests are used to calculate a *p-value*. If a *p-value* for a test is determined to be equal to 1, then the sequence appears to have perfect randomness. A *p-value* of zero indicates that the sequence appears to be completely non-random [16]. A significance level ($\alpha$) can be chosen for the tests. If *p-value* $\geq \alpha$, then the sequence appears to be random. If *p-value* < $\alpha$, then the sequence appears to be non-random. Typically, $\alpha$ is chosen in the range [0.001, 0.01]. An $\alpha = 0.01$ indicates that one would expect 1 sequence in 100 sequences to be rejected. A *p-value* $\geq 0.01$ would mean that the sequence would be considered to be random with a confidence of 99%. A *p-value* < 0.01 would mean that the sequence is non-random with a confidence of 99%.

The *Statistical Test Suite* randomness tests are applied with $\alpha = 0.01$. The randomness results of a PN sequence consisting first 100,000 bits generated by the proposed generator are listed in the Table 1. It is clear that the output under examination has passed all above mentioned randomness tests. The *p-value* obtained in each tests is extensively greater than chosen $\alpha$, which confirms the high randomness of the generated keystream.

Table 1. NIST statistical randomness tests results.

| **Randomness Test** | *p-value* | **Results** |
|---|---|---|
| *Frequency Test* | 0.810072 | *Success* |
| *Block Frequency Test* | 0.610996 | *Success* |
| *Cusum-Forward Test* | 0.355770 | *Success* |
| *Cusum-Reverse Test* | 0.534965 | *Success* |
| *Runs Test* | 0.349349 | *Success* |
| *Longest Runs Test* | 0.735975 | *Success* |
| *Rank Test* | 0.707940 | *Success* |
| *FFT Test* | 0.622387 | *Success* |
| *Linear Complexity Test* | 0.799903 | *Success* |
| *Serial Test* | 0.468962 | *Success* |
| *Approx Entropy Test* | 0.360733 | *Success* |
| *Lempel-Ziv Compression* | 0.993734 | *Success* |
| *Overlapping Template* | 0.561126 | *Success* |

### 3.2 Encryption Results

To evaluate the encryption performance of the proposed keystream generator, it is employed to encrypt the voice data. An original voice signal having 59114 samples, sampled at rate of 16 KHz is encrypted using the keystream generated out of the system. The voice signal is pre-processed and quantized to get the corresponding voice bitstream. The voice bitstream is then XORed with the output keystream. The simulation result of voice encryption is shown in Fig. 3. As it can be seen that the encrypted voice signal shown in Fig. 3(b) is totally distinct from the original voice signal shown in Fig. 3(a) and it is randomly distributed like a noise signal. The signal distribution in Fig. 3(b) is completely flat/uniform at two extreme ends. This shows the effectiveness and suitability of the proposed scheme for voice data encryption.

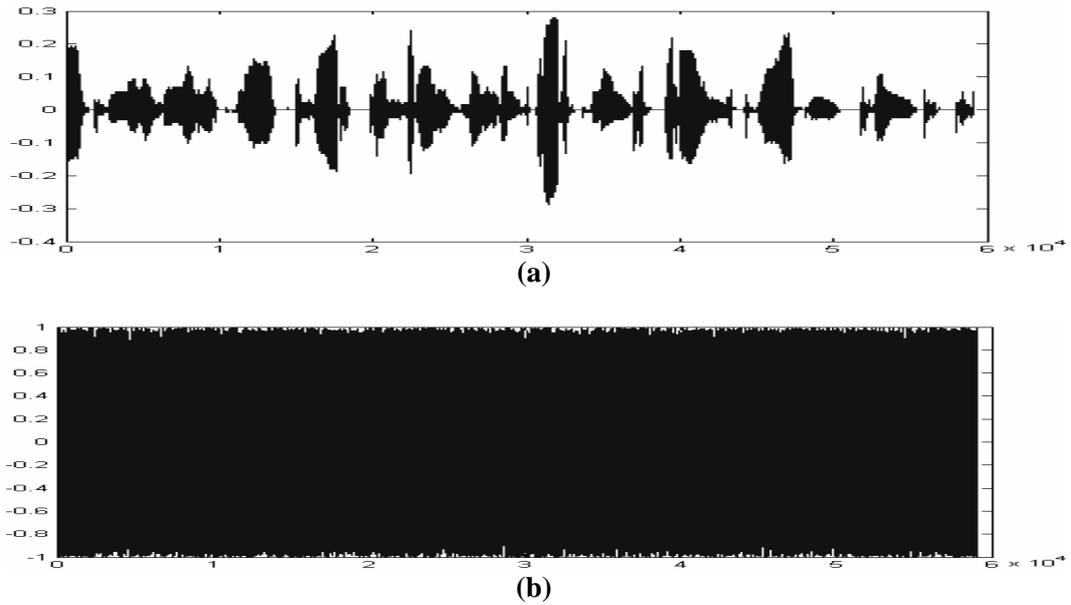

Figure 3. Voice encryption: (a) Original voice signal (b) Encrypted voice signal

### 3.2.1. *Auto-correlation*

The auto-correlation function depicts the random distribution of a signal. According to the Golomb's randomness postulate, a high random sequence should have equality/uniformity in signal distribution and auto-correlation is delta-function. The auto-correlation of the original and encrypted signals are sketched and shown in Fig. 4. It is evident from the plot shown in Fig. 4(b) that the encrypted voice signal has delta-function form. The auto-correlation functions of original and encrypted voice signals have a maximum value of 0.8707092 and 0.0152536 for non-zero shift, respectively. Hence, the encrypted signal is exhibiting a random signal like characteristics.

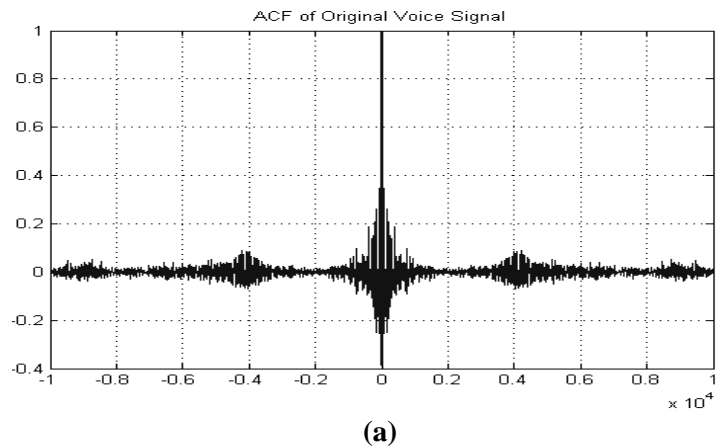

(a)

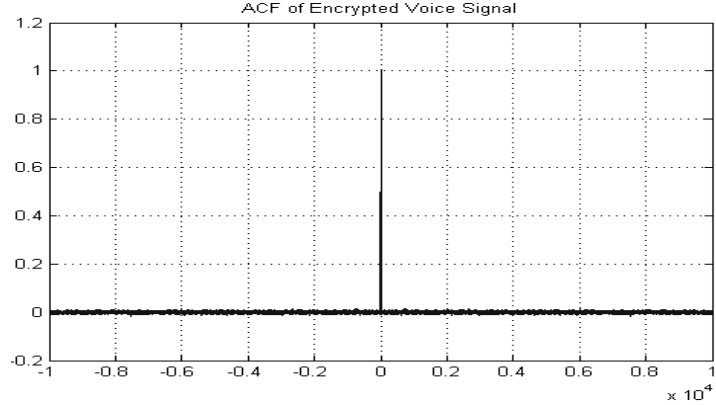

**(b)**

Figure 4. Auto-correlation function of (a) Original and (b) Encrypted Voice Signals

### 3.2.2. Percent Residual Deviation

In order to determine the extent to which the encrypted signal is deviated from the original signal, Sufi *et al.* [17] uses percent residual deviation (PRD) parameter defined in Eq. 8. The parameter provides the measure of dissimilarity between original and encrypted signals. The percent residual deviation ($\psi$) for original $O(i)$ and encrypted $E(i)$ voice signals comes out as 1695.196, where as it is found to be 0.0 for original and decrypted signals. This shows that the encrypted signal is highly deviated from its original signal.

$$\psi = 100 \times \sqrt{\frac{\sum_{i=1}^{n}[O(i)-E(i)]^2}{\sum_{i=1}^{n}O^2(i)}} \quad (8)$$

### 3.2.3. Key Space

The key space of the encryption system should be large enough to resist the brute-force attack. In the proposed scheme, all initial conditions and parameters constitute the secret key of encryption system. For a $10^{-10}$ floating point precision, all key parameters can take $10^{10}$ possible values. Therefore, the key space comes out as $t \times (10^{10})^{12} \approx 2^{408}$, which is large enough to resist the exhaustive attack. The proposed voice encryption system is highly sensitive to a tiny change in secret keys.

### 3.2.4. Key Sensitivity

To demonstrate the key sensitivity, only one parameter of key is changed at a time by a tiny amount of $\Delta = 10^{-10}$, keeping all other parameters of key unchanged and the scheme is applied to recover the voice signal. The results of demonstration are shown in Fig. 5. To quantify the sensitivity, the percentage difference between original and recovered voice signals is calculated and listed in the Table 2. It is clear from the Fig. 5 and Table 2 that voice recovered with tiny changed key has random behaviour and is totally different from the original voice.

Table 2. Percentage difference between original voice and recovered voice.

| # | Test | % Difference |
|---|------|--------------|
| 1 | $x_1(0) + \Delta$ | 99.664 |
| 2 | $x_2(0) + \Delta$ | 99.617 |
| 3 | $x_3(0) + \Delta$ | 99.629 |
| 4 | $\sigma + \Delta$ | 99.599 |
| 5 | $\rho + \Delta$ | 99.583 |
| 6 | $r + \Delta$ | 99.630 |
| 7 | $t + 1$ | 99.572 |
| 8 | $y_1(0) + \Delta$ | 99.657 |
| 9 | $y_2(0) + \Delta$ | 99.648 |
| 10 | $y_3(0) + \Delta$ | 99.639 |
| 11 | $a + \Delta$ | 99.582 |
| 12 | $b + \Delta$ | 99.641 |
| 13 | $c + \Delta$ | 99.576 |
| 14 | $\Delta = 0$ | 0.0 |

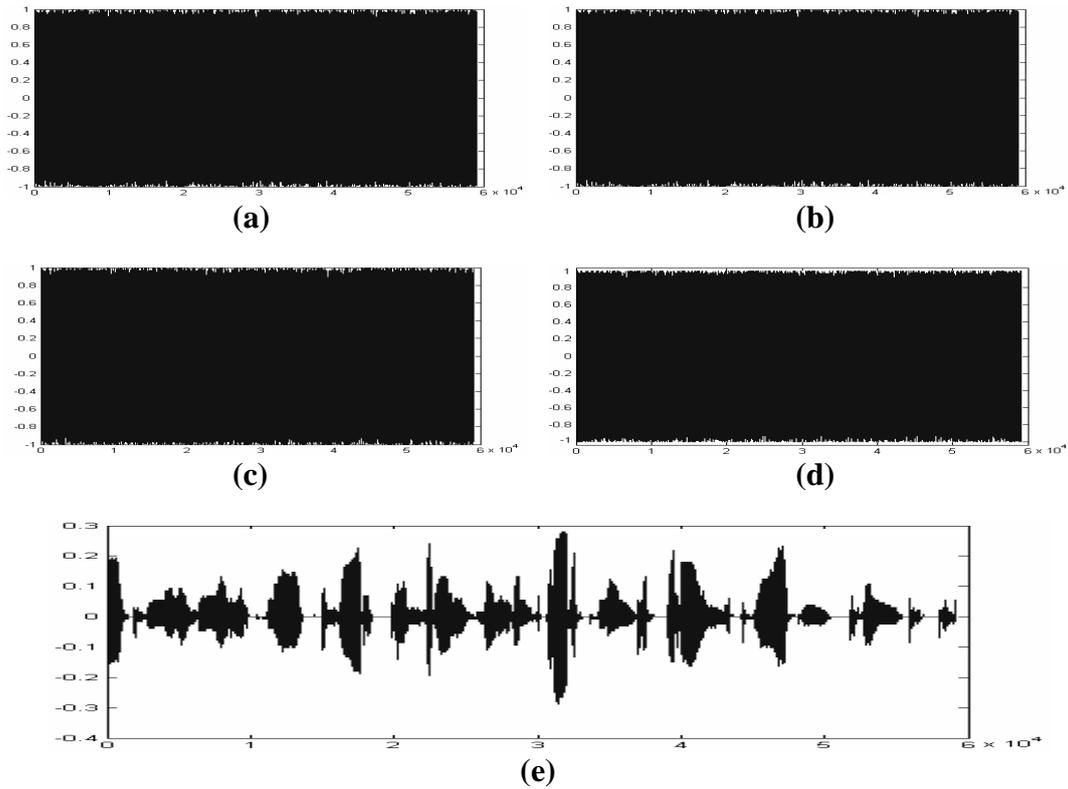

Figure 5. Recovered voice signals with (a) $x_1(0) + \Delta$, (b) $x_3(0) + \Delta$, (c) $y_2(0) + \Delta$, (d) $c + \Delta$ and (e) $\Delta = 0$ i.e. correct key

## 3. CONCLUSIONS

In this paper, a chaos based keystream generator is proposed for voice data encryption. The voice data bitstreams are encrypted using chaotically mixed keystream. High dimensional chaotic systems like Lorenz and Chen are employed to generate more complex and unpredictable six chaotic sequences. After quantization and mixing operations, the system generates statistically and cryptographically better encryption keystream. Experimental analysis demonstrates the generated keystream has high randomness. Moreover, the encryption results also confirm the effectiveness of the scheme for voice data encryption. The results of statistical analyses like randomness, auto-correlation function, signals distribution, percent-residual deviation, key space and key sensitivity indicate high security and suitability of the proposed scheme for practical voice data encryption.